# PROPER-MOTION MEASUREMENTS OF THE CYGNUS EGG NEBULA

Toshiya Ueta[1,2], Koji Murakawa[3], Margaret Meixner[4]

## ABSTRACT

We present the results of proper-motion measurements of the dust shell structure in the Egg Nebula (AFGL 2688, CRL 2688, V1610 Cyg), based on the archived two-epoch imaging-polarimetry data in the $2\mu m$ band taken with NICMOS on-board the *Hubble Space Telescope*. We measured the amount of motions of local structures in the nebula by determining their relative shifts over an interval of 5.5 years. The dynamical age of the nebula is found to be roughly 350 years based on the overall motion of the nebula that exhibits a Hubble-law-esque linear relation between the measured proper motion and the projected radial distance from the origin of the expansion. By adopting the de-projected velocity of 45 km s$^{-1}$ at the tips of the bipolar lobes, our proper-motion measurements indicate that the distance to the Egg Nebula is about 420 pc and that the lobes are inclined at $7\overset{\circ}{.}7$ with respect to the plane of the sky. The refined distance estimate yields the luminosity of the central star of $3.3 \times 10^3$ L$_\odot$, the total shell mass of 1.2 M$_\odot$, and the mass loss rate (the upper limit) of $3.6 \times 10^{-3}$ M$_\odot$ yr$^{-1}$. Assuming 0.6 M$_\odot$ central post-AGB stellar mass, the initial mass of the Egg is 1.8 M$_\odot$. Given the adopted distance to the nebula and the tip velocity, the lobe tips would need only 270 years to reach their epoch 2 positions. This is about 80 years shorter than the dynamical age of the nebula. This discrepancy, together with the fact that the lobe tips do not follow the linear relation as the rest of the nebula, suggests that the tips increased their velocity due to shock acceleration. Upon analysis, we also discovered that the central star of the Egg Nebula has proper motion of its own at the rate of (14, −10) mas yr$^{-1}$ and that the apparent bipolar lobes consist of multiple outflows at distinct inclination angles projected onto each other.

*Subject headings:* circumstellar matter — stars: AGB and post-AGB — stars: mass loss — stars: individual (AFGL 2688, CRL 2688, Egg Nebula, V1610 Cyg) — reflection nebulae

## 1. INTRODUCTION

The Cygnus Egg Nebula (AFGL 2688, CRL 2688, V1610 Cyg; hereafter, the Egg) is one of the most extensively studied post-asymptotic giant branch (post-AGB) stars to date since its discovery as an infrared source accompanied by a reflection nebulosity (Ney et al. 1975). The nebula, usually referred to as a proto-planetary nebula (proto-PN; Kwok 1993; Van Winckel 2003), consists of the post-AGB central star and the circumstellar dust shell that has been ejected from the central star mostly through mass loss during the AGB phase. In proto-PNs, photoionization of the circumstellar matter has not yet been initiated. Therefore, the proto-PNs' circumstellar density structure retains valuable information pertaining to AGB mass loss.

In terms of the nebula structure, the Egg possesses many characteristic traits (Goto et al. 2002 for a recent comprehensive summary), some of which pose serious problems for their coexistence in the nebula due to contradictory formation theories for the respective structures. Although the Egg has been treated as an "archetypical" proto-PN, an amazingly complex structure of the nebula makes this object rather unique.

In order to unravel the mass loss processes, it is essential to understand the fundamental properties of the nebula: these include, for example, the mass of the nebula, the mass loss rate and the dynamical age of the nebula. Since these quantities depend on the distance to the central star, an accurate distance estimate is a prerequisite. However, the distance determination is one of the most difficult tasks in evolved star research, and hence, distance to evolved stars in general remains an elusive quantity.

Proper-motion measurement of the circumstellar nebula provides a direct means to determine the distance to the nebula, assuming that the expansion of the nebula has been constant over the lifetime of the nebula. The technique has been used to determine the distances to PNs based on photographic plates taken over a large ($\sim 50$ years) time interval (e.g. Liller et al. 1966) and high-resolution VLA mapping over an interval of several years (e.g. Masson 1986). After 15 years of operation, the *Hubble Space Telescope* (*HST*) has furnished high-resolution multi-epoch data that can be used in proper-motion measurements (Borkowski & Harrington 2001; Li et al. 2002).

Because of its astonishing nebula structure, the Egg has been a popular recurring target of the *HST*. The Egg has been observed twice in the NICMOS imaging polarimetry mode over a time interval of 5.5 years. Naturally, such a data set is ideal for proper-motion studies. In the following, we present the results of proper-motion measurements of the Egg based on the two-epoch imaging polarimetric data. We first describe the data set and reduction procedure (§2), then discuss the method of proper-motion measurements and the results (§3), and finally summarize conclusions (§4).

## 2. OBSERVATIONS AND DATA REDUCTION

For the present work, we used high-resolution near-IR imaging-polarimetric data of the Egg obtained with *HST*/NICMOS. The Egg has been observed twice in the NICMOS polarimetric mode. The data were taken first on 1997 April 8 (hereafter, epoch 1), as part of the early release ob-servations after the installation of NICMOS in the Servicing Mission 2 during Cycle 7 (SM2/ERO 7115 by D. Hines), and second on 2002 October 6 (hereafter, epoch 2), as part of the calibration observations (CAL/NIC 9644 by D. Hines) after the installation of NICMOS cooling system in the Servicing Mission 3B (SM3B) during Cycle 11. These observations provide a 5.5-year time interval for proper-motion studies. Although the Egg was observed with both of the short (POL-S)[5] and long (POL-L) wavelength polarizers, the smaller field-of-view in the POL-S band failed to capture the nebulosity of the Egg in its entirety. Thus, we used only the POL-L data for the present analysis. Table 1 lists the major observing parameters for the two sets of POL-L data.

[1] NASA Ames Research Center/SOFIA, Mail Stop 211-3, Moffett Field, CA 94035, USA; tueta@sofia.usra.edu
[2] NRC Research Associate/NPP Research Fellow
[3] Max-Planck-Institut für Radioastronomie, Auf dem Hügel 69, G-53121, Bonn, Germany
[4] Space Telescope Science Institute, 3700 San Martin Drive, Baltimore, MD 21218, USA

For data reduction, we used the standard set of NICMOS calibration pipeline programs provided in IRAF/STSDAS Version 3.1[6]. Additional care was taken to remove the Mr. Staypuft (the amplifier ringing and streaking due to bright targets) and the pedestal effect (a variable quadrant bias) anomalies. To remove the Mr. Staypuft, a set of "undopuft" IDL routines (provided by the STScI NICMOS group) was applied to the raw data before using the CALNICA routines. The pedestal effect was removed by adding the *biaseq* and *pedsub* tasks during the otherwise default CALNICA processes.

The anomaly-cleaned, calibrated frames of dithered images were then combined into a single image by employing the variable-pixel linear reconstruction algorithm (the STSDAS package *dither*, Version 2.0; Koekemoer et al. 2002). The NICMOS arrays are slightly tilted with respect to the focal plane, and thus the NICMOS pixel scales along the two spatial axes of the data frame are not identical. When the dithered frames are drizzled, we rectified all data frames using the distortion coefficients determined from the plate scale of the date of observations (Table 2). Upon distortion correction, all the data were rectified and rescaled to have $0.''07610$, which is the geometric mean of the epoch 1 plate scales. In addition, the distortion-corrected data were sub-pixelized by a factor of four when drizzled. The final images are therefore in $0.''01903$ pixel$^{-1}$. Furthermore, images are de-rotated upon drizzling so that the resulting images have the nominal image orientation (north is up, east to the left).

The rectified, drizzled images for each polarizer were then deconvolved by the *lucy* task (with 30 iterations) using the PSF data constructed by the TinyTim software (Version 6.3) available from STScI. At this stage, there are in total six deconvolved images, one for each of the three polarizers (0, 120, and 240 degrees) for two epochs. Before obtaining the Stokes images for each epoch, we must align three polarizer images. To do so, we identified as many as seven background stars in the field and determined shifts among these three images by employing a cross-correlation method on those background stars. All image frames were aligned using the 0 degree polarizer image of epoch 1 as the reference frame. The shifts among the image frames of the same epoch were found to be at most 4.9 pixels ($0.''093$), while the shifts among the image frames of the different epochs were at most 8.7 pixels ($0.''17$). The alignment error was determined to be 0.4 pixels ($0.''0082$) and 0.3 pixels ($0.''0057$) respectively along the x and y directions in the final images. To shift and align the images, we used bicubic spline interpolation.

The aligned frames were photometric-calibrated before the Stokes images were derived through the matrix inversion method (Hines et al. 2000; Dickinson et al. 2002). Since the detector operating temperature is different before and after SM3B, we have two sets of the photometric conversion factors and matrix coefficients. For photometric calibration, we adopted the conversion factors that we determined in our previous analysis of the NICMOS imaging polarimetric data, while for matrix conversion we used the coefficients provided by the STScI NICMOS group. These values are listed in Table 2 of Ueta et al. (2005a), in which NICMOS imaging-polarimetric data reduction is discussed in detail.

## 3. PROPER-MOTION MEASUREMENTS

After completing the reduction steps described above, we obtained a set of two-epoch Stokes images for the Egg. In the subsequent analysis, we used only the Stokes $I$ (the total flux) images in our proper-motion measurements. The polarimetric properties of the Egg are discussed elsewhere (Sahai et al. 1998; Weintraub et al. 2000; Ueta et al. 2005b). To determine proper motion of the nebulosity over the 5.5-year time interval defined by these two-epoch observations, we followed the method elucidated by Currie et al. (1996) and Morse et al. (2001) in their proper-motion measurements of $\eta$ Car. This method is based on the idea that proper motion of any local structure in a nebula due to an expansion can be measured by determining the amount of translational shift of the structure via a cross-correlation analysis. The following analysis was done in IDL using scripts we created making use of the IDL Astronomy User's Library.[7]

We identified 42 distinct local structures distributed throughout the nebula (14 in each of the north and south lobes and seven in each of the east and west arms) in the epoch 1 image. These structures should be located at slightly shifted positions in the epoch 2 image due to proper motion of the nebula, i.e. a shell expansion. To determine the amount of shift, we first defined square image sections centered at each of these local structures in the epoch 1 image. The size of these image sections has to be large enough so that the segmented structures can be uniquely identified. Meanwhile, the size of the image sections has to be small enough so that the cross-correlation algorithm (to be outlined below) would recognize these structures. Then, we set up, in the epoch 2 image, a search box that is three times larger than the image section centered at the epoch 1 position of the local structures.

In each search box, we took an image section that is the same size as the epoch 1 image section and compared the two image sections taken from the epoch 1 and 2 data by means of cross-correlation. We did this by computing the inverse of the sum of the squares of the difference between the two values in the same pixel in the two image sections. Since this value represents the "correlation" between the two image sections, it tends to be large if the two image sections capture the same local structure (hence, the two are similar) and small if the two are very different. We repeated this operation while the epoch 2 image section was moved around inside the search box pixel by pixel. When this operation was completed, we were left with an array of values, each of which is a measure of correlation between the image sections. Finally, we obtained the amount of translational shift for each of the selected local structures by locating the position of the maximum cor-relation measure. The derived relative shifts of the structures are proper motions of these local structures of the shell over a time interval of 5.5 years.

---

[5] The first epoch POL-S observations were done on 1998 April 9 as GO 7423 by R. Sahai.

[6] IRAF is distributed by the NOAO, which is operated by the AURA, under cooperative agreement with the NSF. STSDAS is a product of the STScI, which is operated by AURA for NASA

[7] The IDL Astronomy User's Library is publicly available through http://idlastro.gsfc.nasa.gov/.

The amount of shift of the structures turned out to depend on the size of the square image section used to define the local structure segments. Hence, we used 11 different sizes from $0.''5$ to $1.''5$ with an interval of $0.''1$. The smallest image section size of $0.''5$ is adopted because the structure elements could not be uniquely identified if the size was smaller than $0.''5$. The largest size of $1.''5$ is chosen because all 42 structure elements used in this analysis is sufficiently smaller than $1.''5$, that is, the use of any larger size would not correctly detect any shift (i.e. most of the structure in the image segment would be stationary, and hence, any local structures in the image segment are mistakenly thought to have been stationary, too). We then took the average of the resulting shifts and adopted the average as the shift of the particular local structure. The quoted error of the shift below is therefore the standard deviation of the shifts obtained by the use of differently sized image sections.

### 3.1. Inferred Proper Motion of the Central Star

Presumably, the shell expansion is caused by a generally radially outward flow emanating from the center of the shell. This has been repeatedly corroborated by past observations using various molecular species as tracers (e.g. Bieging & Nguyen-Q-Rieu 1988, 1996; Yamamura et al. 1996; Cox et al. 2000; Lim & Trung 2004). Therefore, the aforementioned analysis is expected to reveal the selected local structures having shifted radially outward with respect to some central location in the nebula over the past 5.5 years.

However, our analysis indicated that the *entire* nebula has been shifted towards the northeast direction, except for the very tip of the south lobe. Figure 1 displays the derived proper motion of the local structures as vectors overlaid with the Stokes *I* image. Each vector, located at the position of the corresponding local structure, expresses the direction and amount of the shift. The maximum and minimum shifts are respectively $10.8 \pm 0.1$ and $1.3 \pm 0.9$ pixels. Using these vectors, we can backtrack the origin of the outflow. This is done by searching for a location that minimizes the sum of the squares of the normal distance to each vector. The origin of the outflow was found to be located outside the nebula (indicated by the plus sign in Figure 1), which is $(-4.''8, -3.''8)$ away from the isolated intensity peak at the southern tip of the north lobe (Peak A, which is also indicated in Figure 1; Weintraub et al. 2000).

This translation behavior of the shell structure is clearly odd, and there seem to be three possibilities as the cause. The first possibility is that the image rectification and alignment were not optimally done with the background stars. If the image distortion still remained in the final images it should have been very difficult to align the two-epoch data, because the position of the stars should have been shifted out of sync and some stars could have moved more than others, leading to a range of alignment errors for each star. However, we do not find such variable alignment errors, and the alignment error was found to be a fraction of a pixel (at $0.''01903$ pixel$^{-1}$) for all background stars used in the alignment analysis. Furthermore, it is inconceivable that the background stars keep their relative positions the same between the two-epoch data while the intervening pixels still possess appreciable distortions. Hence, we conclude that this possibility is unlikely.

The second possibility is that the background stars have shifted their positions due to their own proper motions. However, the alignment error was found to be very small, i.e. the background stars maintained their relative positions. Therefore, if this is the case all background stars must have shifted their positions by the same amount in the same direction. This is very unlikely. The third possibility is that the Egg itself has shifted its position in the sky due to its own proper motion. If this is the case, it makes very natural sense that all background stars appear to have been shifted in the same way. Thus, we concluded that this apparently strange translation behavior of the nebula derived from the background-star-aligned two-epoch data was due to proper motion of the Egg itself.

This apparent translation then represents proper motion of the central star, provided that the nebula and central star share the same bulk motion. By applying the derived shifts between the epochs, we can remove the bulk motion of the nebula and put the subsequent nebula motions into the frame of reference of the central source. We now need to determine proper motion of the Egg itself over the past 5.5 years before we can perform proper-motion study of the nebula. Proper motion of the Egg itself should be less than about 11 pixels, which is the maximum local structure shift found in the background-star-aligned two-epoch data. To determine proper motion of the Egg, we first shifted the epoch 2 image pixel by pixel for $\pm 10$ pixels in the x and y directions. At each of the shifted epoch 2 image position, we measured the amount of translation for all 42 local structures in the same way as explained above. The spatial translation of the local structures should become radially outward, if the two epoch images are optimally aligned. Then, the sum of the shifts for the local structures, considering both the *amount* and *direction* of the shifts, should be minimal. Thus, we determined the optimum epoch 2 image shift by identifying the epoch 2 image location with respect to the epoch 1 image that would minimize the sum of the translation vectors of the local structures. Proper motion of the Egg itself is then equal to the amount of the optimum epoch 2 image shift.

However, as we have already seen, the length and direction of the translation vectors of the local structures would depend on the size of the image sections used in the analysis. Thus, we repeated this analysis for different image section sizes between $0.''5$ and $1.''5$, and adopted the averaged epoch 2 image shift as the optimum value. The required shift to the epoch 2 data was thus found to be $(4.0 \pm 0.6, -2.9 \pm 0.6)$ in pixels. Again, the quoted errors are the standard deviation of the shifts obtained for different image section sizes. By converting these values in mas yr$^{-1}$, we find that proper motion of central star of the Egg is $(13.7 \pm 2.0, -10.2 \pm 2.0)$ mas yr$^{-1}$.

### 3.2. Differential Proper Motion in the Nebula

Figure 2 shows the derived proper motion of the nebula determined based on the epoch 2 image that is corrected for the stellar proper motion. The vectors now show a highly radially symmetric pattern, and the origin of the shell expansion is found to be located in the central region of the nebula (depicted by the larger plus sign in Figure 2), which is $(0.''7, -0.''5)$ away from Peak A. The maximum and minimum (relative) shifts of the nebula structures are respectively $7.7 \pm 0.3$ and $1.4 \pm 0.1$ pixels. The expansion origin is indeed close to the center of the polarization vector pattern (depicted by the smaller plus sign in Figure 2; Weintraub et al. 2000) and the origin of multiple collimated CO outflows (Cox et al. 2000). In retrospect, this consistency increases our confidence in the above analysis in concluding that the Egg has experienced proper motion of its own over the past 5.5 years.

Figure 3 displays ratio images (epoch 2 over epoch 1) in the Stokes $I$ (left panel) and in polarized light, $I_{\rm pol}$ ($= \sqrt{Q^2+U^2}$; right panel). Since the POL-L filter covers $H_2$ v=1-0 S(1) emission at 2.12$\mu$m and such has been detected from the Egg (Beckwith et al. 1984; Latter et al. 1993; Hora & Latter 1994; Cox et al. 1997; Sahai et al. 1998; Kastner et al. 2001), the Stokes $I$ ratio image does not necessarily show only the motion of matter in the nebula. For example, a sudden increase in $H_2$ line emission in the epoch 2 image can be confused as a shift of some local structure. The $I_{\rm pol}$ images, on the other hand, recover only scattered (i.e. polarized) light in the nebula, and therefore, trace only the presence of scattering matter (e.g. Ueta et al. 2005a,b). Hence, the $I_{\rm pol}$ ratio image shows shifts of local structures, via scattering matter, without any confusion by $H_2$ line emission.

By inspection, both of the Stokes $I$ and $I_{\rm pol}$ ratio images appear similar, except for the equatorial region. This indicates that the Stokes $I$ maps trace the motion of scattering matter in the lobes quite well and that proper motion of the nebula can be measured from them. However, this is not the case in the equatorial region, i.e. in the east and west arm structures. There is very little $I_{\rm pol}$ detection at the tips of the east and west arms because (1) most of the structure in the east-west direction is due to $H_2$ line emission (e.g. Cox et al. 1997 and references therein) and (2) scattered photons are absent due to heavy extinction in the dense equatorial region (e.g. Ney et al. 1975; Sahai et al. 1998). Thus, the observed shift in the surface brightness distribution in the east-west arms can be affected by the sweeping of the $H_2$ exciting shock front and may not be appropriate for determining the material motion. Therefore, we will have to follow the shift in the lobes away from the equatorial region.

In CO J=2-1 interferometer mapping, Cox et al. (2000) found complex velocity structures in the nebula, in which velocities increase as the distance from the center increases along the north-south lobes, while the highest velocities are found closer to the center along the east-west arms. In $H_2$, Kastner et al. (2001) found fast moving material close to the center in spectral imaging. Our proper-motion measurements are consistent with these previous reports. Figure 4 shows plots of measured proper motions (converted to arcsec yr$^{-1}$) as a function of projected radial distance from the origin of the expansion determined from about 85 independent positions in the north-south lobes and the east-west arms. Material motions in the north-south lobes show the velocity gradient in which the fastest matter is located at the farthest from the origin, while in the east-west arms the gradient is reversed. Thus, the two-epoch Stokes $I$ data show bulk material motion along the north-south lobes and motion of the shocked matter ($H_2$) along the east-west arms. In general, we found larger errors along the east-west direction. This is indicative that the apparent shifts of local structures in the east-west direction are due to changes in emission structure caused by shocked $H_2$ emission (whose structure is not necessarily preserved over the two epochs, and hence, it is more difficult to find any "correlation") and is not due to bulk material motion (which tends to maintain local structure in translation, and hence, there is a stronger correlation).

The apparent Hubble-law-esque expansion of the north and south lobes seen in our measurements (Figures 2 and 4) maintain the overall shape of the lobes as they expand. Also shown in Figure 4 are the least-squares fits to the data points. The solid black line and solid light gray line correspond to respectively the fits to the measurements in the north and south lobes, while the dashed dark gray line is the fit to all measurements in the two lobes. Assuming there is no acceleration in material motion in the lobes, the reciprocal of the slope of the fitted line is equivalent to the elapsed time since the beginning of the expansion, i.e. the dynamical age of the nebula. From the fit, we find $345 \pm 2$, $393 \pm 5$, and $352 \pm 1$ years respectively for the north lobe, the south lobe and the entire north-south lobes. Thus, based on the overall expansion of the lobes, the dynamical age of the nebula is roughly 350 years.

If the bipolar structure of the Egg is generated by the superwind (Renzini 1981; Iben & Renzini 1983), as suggested by the previous imaging surveys of the innermost region of the post-AGB shells (Meixner et al. 1999; Ueta et al. 2000), the superwind in the central star of the Egg began emanating about 350 to 400 years ago in early to mid 1600's. This finding places the upper limit to the duration of the post-AGB evolution of the central star. The central star of the Egg has become of F5I spectral type since the star left the tip of the AGB, which corresponds to the end of the appreciable AGB mass loss. Our measurement suggests that it has been at most 350 years. According to the evolution models of Blöcker (1995), the initial mass of such a star has to be less than about 3 M$_\odot$.

### 3.3. *Distance*

We can determine the distance to the Egg using our proper-motion measurements, if we know the sky-projected outflow velocity, $v_{\rm proj}$, of the corresponding local structures. However, observational velocity measurements are sensitive only to the line-of-sight outflow velocity, $v_{\rm los}$, which is the velocity projected along the line of sight[8]. Thus, we have to deduce $v_{\rm proj}$ from $v_{\rm los}$ using the inclination angle of the bipolar axis with respect to the plane of the sky, $i$. If we express the time interval for the two-epoch observations by $t$ (in years) and the measured proper motion of a particular nebula structure by $\theta$ (in arcsec), for which $v_{\rm proj}$ (and/or $v_{\rm los}$ in km s$^{-1}$) is estimated by other means, we can compute the distance, $D$ (in pc), to the Egg by

$$D = 0.211 \frac{v_{\rm proj} t}{\theta} = 0.211 \frac{v_{\rm los} t \cot i}{\theta} \quad (1)$$

since $v_{\rm proj} = v_{\rm los} \cot i$.

Cox et al. (2000) measured the line-of-sight velocity of $6 \pm 0.8$ km s$^{-1}$ in CO J=2-1[9] at the tips of the north-south lobes with respect to the systemic velocity of the nebula (positions F1 and F2 in their designation, which is indicated in Figure 3). For structures around F1 and F2, we obtain $\theta = 0''.123 \pm 0''.002$. The time interval for the present two-epoch observations is 5.498 years. Hence, we need to constrain $i$ to determine the distance to the Egg.

Past molecular observations revealed the velocity structure of the expanding molecular shell around the Egg in detail. Young et al. (1992) and Yamamura et al. (1996) detected multiple outflow components of fast ($\sim 40$ km s$^{-1}$) and slow ($\sim 20$ km s$^{-1}$) winds in the Egg. Based on their model calculations to fit various line shapes, Young et al. (1992) argued that the higher velocity component (of $\sim 40$ km s$^{-1}$) is located within $1.5 \times 10^{17}$ cm of the central star while the lower velocity component (of $\sim 20$ km s$^{-1}$) surrounds the fast component and extends up to roughly twice farther than the fast component. Cox et al. (1997) compared the observed H$_2$ line ratios and intensities with shock excitation models and concluded that H$_2$ emission from the Egg is fully compatible with shock excitation in relatively dense gas ($10^6$ cm$^{-3}$) and shock velocities of about 20–40 km s$^{-1}$. In fact, Cox et al. (2000) successfully resolved outflow structures in the inner $\sim 16'' \times 16''$ region of the nebula with the $\sim 1''$ beam size, showing a collimated outflow at $\sim 45$ km s$^{-1}$ nearly aligned with the line of sight (at G1 and G2 positions in Figure 3; also see Figure 3 of Cox et al. 2000). The position-velocity maps of Cox et al. (2000) also show collimated outflows into the F1/F2 and D1/D2 directions. Assuming these collimated outflows have the same velocity based on the work by Young et al. (1992) and Cox et al. (2000), we adopt 45 km s$^{-1}$ (with 20% measurement error) as the de-projected "radial" outflow velocity at the tips of the lobes. In order for the radial (de-projected) outflow velocity of 45 km s$^{-1}$ at F1/F2 to have the line-of-sight velocity of 6.0 km s$^{-1}$, the outflow has to be inclined at $7.°7$ with respect to the plane of the sky. Thus, using $v_{los} = 45$ km s$^{-1}$, $t = 5.498$ yr, $\theta = 0.''123$, and $i = 7.°7$, we obtain $420 \pm 60$ pc for the distance to the Egg.

---

[8] The line-of-sight velocity is usually referred to as the radial velocity (in the heliocentric sense). However, we have used the word "radial" to describe the radial outflow of the circumstellar nebula (in the stellarcentric sense). Thus, we use the term "line-of-sight velocity" instead of the more common "radial velocity" in this paper to make clear distinction.

[9] 0.8 km s$^{-1}$ is the effective velocity resolution of the measurements in Cox et al. (2000).

At this distance, the high velocity wind component seen by Young et al. (1992) should extend as far out as $24''$. Since the bipolar lobes are elongated about $7''$, the lobes are well within the region in which the higher velocity component wind is found. Moreover, Lopez & Perrin (2000) concluded that the inclination angle of the Egg is less than 10° by performing radiative transfer modeling in a large parameter space with an assumption that the shell density distribution function is radially and latitudinally decreasing and a requirement that models fit to the surface brightness distribution from 0.55 to 18 $\mu$m. Therefore, we conclude that the distance to the Egg is $420 \pm 60$ pc, provided that the de-projected lobe tip velocity is $45 \pm 9$ km s$^{-1}$ and the inclination angle of the bipolar lobes is $7.°7 \pm 1.°9$.

This distance estimate of 420 pc is a factor of two to three smaller than the value of $1-1.5$ kpc that has been typically adopted in the past. While the previous distance estimates were based on various arguments, the most quantitative argument is probably the one presented by Cohen & Kuhi (1977), in which the interstellar extinction of $A_V = 4.0 \pm 1.8$ was determined based on their measurements of photospheric Na I D lines reflected in the nebula lobes and the distance of $\sim 1.5$ kpc was suggested using the extinction measurements by FitzGerald (1968). Recently Schlegel et al. (1998) have presented a new Galactic extinction estimator based on the 100 $\mu$m dust IR emission maps constructed with COBE/DIRBE and IRAS/ISSA data. This estimator yields E(B-V) of 0.28 towards the direction to the Egg, which is equivalent to $A_V = 0.88$ assuming $R_V = 3.1$. Therefore, this new $A_V$ estimate suggests that the distance to the Egg is closer than $1-1.5$ kpc.

Yusef-Zadeh et al. (1984) derived the inclination angle of 16° based on multiple-scattering Monte Carlo radiative transfer calculations to fit the surface brightness distribution in the optical with a radially and latitudinally decreasing density distribution. Assuming roughly 20% error in the inclination angle determination, we get $197 \pm 29$ pc as the distance to the Egg. In this case, the de-projected outflow velocity at the tips of the bipolar lobes is 22 km s$^{-1}$, and the dynamical age of the lobe tips is roughly 260 years. However, as mentioned above, the past molecular studies indicate that the Egg is located as far as 1.4 kpc if the tips of the lobes expand at about 20 km s$^{-1}$ (Young et al. 1992). Thus, we do not adopt this alternate inclination angle due to this inconsistency.

Young et al. (1992) also reported detection of a very high velocity wind component at nearly 100 km s$^{-1}$ in CO J=4-3. However, Cox et al. (2000) did not detect such high velocities in the inner $\sim 16'' \times 16''$ region of the nebula in CO J=2-1. Again, shock velocities of about 20–40 km s$^{-1}$ were suggested for the tips of the H$_2$ emission regions from H$_2$ line observations and comparison with shock models (Cox et al. 1997). If we use 100 km s$^{-1}$ as the de-projected outflow velocity of the bipolar tips, we obtain the inclination of $3.°4 \pm 0.°8$ and the distance of $940 \pm 130$ pc. While this distance is consistent with the previous value of $1-1.5$ kpc, we conclude that this case is less likely than our adopted case of $420 \pm 60$ pc based on (1) lack of substantial evidence to support the de-projected outflow velocity of 100 km s$^{-1}$ at the lobe tips and (2) the inclination angle that is too shallow given the model predictions by Lopez & Perrin (2000) mentioned above.

By adopting the distance of 420 pc and the de-projected lobe tip velocity of 45 km s$^{-1}$ (and assuming the lobe tips neither accelerated nor decelerated during the expansion), we find that the time needed for the lobe tips to reach their epoch 2 positions from the center of the expansion is about 270 years. This is about 80 years shorter than the dynamical age of the nebula obtained by fitting the Hubble-law-esque linear relation observed in our proper-motion measurements. This indicates that the tip velocity (45 km s$^{-1}$) is somewhat faster than the velocity of the overall lobe motion. This possible increase of the tip velocity is actually seen in the top frame of Figure 4, in which the projected motions of the local structures near the lobe tips (located at $\gtrsim 5''$) slightly deviate from the Hubble-law-esque linear relation. The fact that the projected motions of local structures within $\sim 5''$ can be fitted very well with a line going through the origin of the plot suggests that the tips increased their velocity with respect to the velocity of the overall lobe motion. Assuming the superwind origin of the bipolar structure of the Egg (i.e. the superwind impinging upon the pre-existing, slowly expanding material ejected by AGB mass loss), the increasing tip velocity suggests that the lobe tips are shock accelerated due to ram pressure from the superwind. The acceleration effect is especially seen well at the tips because of the geometry (small inclination) of the lobes.

Using the bolometric flux estimate of $6 \times 10^{-7}$ erg s$^{-1}$ cm$^{-2}$ (Knapp et al. 1994), the luminosity of the central star of the Egg is found to be $3.3 \times 10^3$ L$_\odot$. This is consistent with the evolution models of $\lesssim 3$ M$_\odot$ initial mass stars (Blöcker 1995). Using 420 pc in the distance-dependent quantities that were derived by radiative transfer modeling (Lopez & Perrin 2000), we find that the total mass of the shell is 1.2 M$_\odot$. Assuming 0.6 M$_\odot$ for the mass of the central star, the initial mass of the star is estimated to be 1.8 M$_\odot$. Similarly, the mass loss rate is found to be $3.6 \times 10^{-3}$ M$_\odot$ yr$^{-1}$. This rate, however, should be considered an upper limit, since the derived value is the maximum in the equatorial plane (in which density is the highest).

### 3.4. *Outflow Structure in the Nebula*

The multiple CO outflow structure in the nebula is demonstrated well in a position-velocity diagram (Figure 3 of Cox et al. 2000). We can employ a similar argument as in the previous section for each of the outflow tips to compute the inclination of the outflows. That is, we solve for the inclination angle for each outflow given the distance of 420 pc and the de-projected velocity of the outflows of 45 km s$^{-1}$ (provided that the lobe tips are all moving at this velocity). While the main lobes (F1/F2) are inclined with $\sim 8°$ (at P.A. = 12°/18°), we find that other outflows along the main lobes (D1/D2 and E1/E2) have respectively 57°/55° (at P.A. = 21°/26°) and 75°/56° (at P.A. = 28°/43°). While the F1/F2 and D1/D2 pairs seem to be aligned, the inclination/P.A. of the E1/E2 pair changes widely. This may indicate the presence of two independent flows in these directions.

The same analysis can also be used to determine the outflow structure in the east and west arms. The flows corresponding to A1, B1, and C1 are respectively inclined with 53°, 14°, and 5° (approaching) at P.A. of 117°, 91°, and 77° at the (de-projected) velocities of 48, 33, and 37 km s$^{-1}$. The B2 flow has 24° inclination (receding) at 266° P.A. at 35 km s$^{-1}$. The inclination and velocity are indeterminable at A2 and C2 because there is an emission enhancement in H$_2$ at A2, which prevents the proper-motion measurements, and there is too little signal to reliably measure proper motion of the local structure at C2. B1 and B2 seem to be co-planer flows. However, it is inconclusive if A1/A2 and C1/C2 pairs are co-planer. We are not able to measure proper motion at G1/G2 due to lack of sufficient signal.

There is also an enhancement in H$_2$ at A$'$. Geometrically, this local structure appears to be another shocked region in the upstream of the A1 outflow. However, if we solve for the inclination of the flow at A$'$ by adopting 420 pc and the de-projected tip velocity of 40 km s$^{-1}$ (the average of A1/B1/C1 de-projected tip velocities) we obtain the angle of 40°. Hence, the A1 and A$'$ outflows appear to be distinct flows. Therefore, the Egg seems to have been shaped by a number of well-collimated outflows that are channeled into only a few P.A.'s, to the direction of the north-south bipolar lobes and the direction of the east-west shocked H$_2$ regions. However, outflows along each P.A. angles are not necessarily aligned at the same inclination.

The "standard model" of the Egg assumes the presence of a single bipolar axis that is orthogonal to the dense equatorial region in which H$_2$ line emission is detected (e.g. Yusef-Zadeh et al. 1984; Latter et al. 1993; Lopez & Perrin 2000). As shown above, H$_2$ line emission appears at the head of collimated outflows into multiple directions along the equator. Although it is inconclusive that these flows are parts of a general equatorial rotation (e.g. Kastner et al. 2001), it is not inconceivable that there are multiple outflows on the equatorial plane. For example, Murakawa et al. (2005) have suggested that some of the multiple lobes seen in the shell of an AGB star, IRC +10 216 (considered less evolved than the Egg on a similar evolutionary path), are actually due to scattered star light that has leaked out of the central dust torus through fissures *in* the torus.

The possible presence of multiple outflows in the N-S directions, however, may not be consistent with the standard model in which one outflow is expected along the bipolar axis. Skinner et al. (1997) presented an alternative model in which the bipolar axis of the nebula lies at P.A. of $\sim 60°$ at $\sim 30°$ inclination angle. Then, the structures seen in the N and E (S and W) are parts of the biconical outflows into the NE (SW) direction. This model permits multiple outflows projected onto a single lobe at P.A. of $\sim 15°$ (which represents the limb of the biconical flow). Our proper-motion measurements do not verify/refute either of the standard or biconical outflow models. However, it is of particular interest to note that there is an expanding shell-like structure in the central 2$''$ of the nebula at P.A. of $\sim 54°$ detected in CO J=2-1 Cox et al. (2000). Goto et al. (2002) also showed that the equatorial dust lane in the L$'$ and M$'$ bands is orthogonal to P.A. of $\sim 50°$. Unfortunately, we can neither confirm nor deny the presence of outflows along this direction, since we were unable to make any proper-motion measurements in this heavily extincted region.

### 3.5. *Concentric Arcs*

One of the remarkable morphological characteristics of the Egg is the presence of concentric arcs. However, their origin is still not understood (Sahai et al. 1998; Hrivnak et al. 2001). With our distance estimate, we can constrain the time scale for the occurrence of the arcs.

Unfortunately, we can not measure proper motion due to lack of discernible local structures on these arcs. Thus, we are not able to directly measure the projected velocity of the arcs. Since the arcs in general appear well beyond the region of the bipolar lobes, i.e. the higher velocity component wind region, we adopt the lower wind velocity of 23 km s$^{-1}$ as the velocity of the arcs (Young et al. 1992). The angular separation of the arcs (0$''$3 – 1$''$1; Hrivnak et al. 2001) therefore translates to 30 – 100 years (10 – 50 years if 45 km s$^{-1}$ arc velocity is assumed).

The problem of these arcs has been that we do not know any physical mechanisms which repeat on the time scale of the arcs. It is inconsistent with either of the pulsation ($\sim 1$ yr) or thermal pulsing ($\sim 10^4 – 10^5$ yr) time scale associated with the AGB evolution. The time scale of $\lesssim 100$ yr is still a factor of 10 or more longer than the pulsation time scale of $\sim 2$ M$_\odot$ stars ($\sim 3$ yr; e.g. Vassiliadis & Wood 1993). Thus, we require a yet unidentified mechanism that will generate these concentric arcs in a time interval of 30-100 years.

### 3.6. *Peak A Structure*

The nature of Peak A at the southern end of the north lobe (Figure 1) has been a mystery. Weintraub et al. (2000) claimed, based on the absence of polarization at the peak, that the peak is self-luminous, and hence, is a companion star to the Egg. On the other hand, Goto et al. (2002) concluded that the peak is a concentration of dust based on their spatially resolved infrared spectroscopy.

The presence/absence of scattering matter at Peak A can be easily checked in the $I_{pol}$ data. In the epoch 1 data, there is little $I_{pol}$ emission at Peak. This is probably why Weintraub et al. (2000) concluded that polarization is absent at Peak A. However, $I_{pol}$ surface brightness has increased rather dramatically in the epoch 2 data, which results in the much-enhanced Peak A structure appearing in the ratio image (right frame in Figure 3). The apparent ring-like structure of Peak A in the $I_{pol}$ ratio image suggests that Peak A has expanded while its brightness has increased. Moreover, proper-motion measurements of Peak A indicate that the peak moves at roughly 8 km s$^{-1}$ into P.A. of 94°. This is consistent with other outflows along the east arm. Thus, we suggest that Peak A is not a self-luminous object such as a companion object and is simply a concentration of scattering material.

## 4. CONCLUSIONS

We have performed proper-motion measurements of the Cygnus Egg Nebula using its archived data of *HST*/NICMOS

imaging-polarimetry over two epochs (with a 5.5-year interval). Measurements are done by determining translational shifts of image sections of size $0\rlap{.}''5$ to $1\rlap{.}''5$ via a cross-correlation analysis.

Through the measurements, we have found the following:

1. After calibration, the epoch 2 image had to be shifted by $(4.0 \pm 0.6, -2.9 \pm 0.6)$ pixels to trace a radially symmetric expansion of the shell in the measured proper motion. This means that the Egg (the central star and the nebula altogether) has experienced proper motion of its own at the rate of $(13.7 \pm 2.0, -10.2 \pm 2.0)$ mas yr$^{-1}$ over the past 5.5 years, provided that the nebula and central star share the same bulk motion.

2. There is a Hubble-law-esque linear relationship in the north-south lobes between the measured proper motion and the projected radial distance from the origin of the expansion. The reciprocal of the slope for the fitted line corresponds to the dynamical age of the nebula based on the overall motion of the lobes, which is found to be about 350 years.

3. By adopting 45 km s$^{-1}$ de-projected outflow velocity at the tips of the north-south lobes from high resolution CO maps, the inclination angle of the lobes is $7\rlap{.}°7 \pm 1\rlap{.}°9$ and the distance to the Egg is $420 \pm 60$ pc.

4. At 420 pc and 45 km s$^{-1}$, the lobe tips would need only 270 years to reach their epoch 2 positions from the origin of the expansion. This is about 80 years shorter than the dynamical age of the nebula (item 2 above). The discrepancy between the two values, together with the slight deviation from the Hubble-law-esque relation seen for the material near the lobe tips (Figure 4), suggests that the lobe tip velocity has increased due to shock acceleration, which is caused by the superwind impinging upon the pre-existing, slowly expanding material ejected by AGB mass loss.

5. At 420 pc, the stellar luminosity is found to be $3.3 \times 10^3$ L$_\odot$ and the total mass of the shell is estimated to be 1.2 M$_\odot$. Assuming 0.6 M$_\odot$ as the mass of the central post-AGB star, the initial stellar mass of the Egg is 1.8 M$_\odot$. These values are very much consistent with the evolution models of $\lesssim 3$ M$_\odot$ star (Blöcker 1995). The upper limit for the mass loss rate is estimated to be $3.6 \times 10^{-3}$ M$_\odot$ yr$^{-1}$.

6. Although the apparent morphology of the nebula is a bipolar nebula with the east-west arm structures, the outflow structure of the nebula is more complex. There are at least three distinct outflows that are projected onto the north-south lobes, while there are at least four distinct, non-co-planer outflows in the east arm.

7. At 420 pc, the concentric arcs seen in the Egg are expected to be associated with outflows of 23 km s$^{-1}$. Then, the spacing between the arcs indicates that the arcs are caused by some mechanisms that works on the time scales of $30 - 100$ years. This corresponds to neither the pulsation nor thermal pulsing of the central star. Another mechanism is needed to explain these arcs.

8. The enigmatic Peak A structure is a concentration of dust and not a self-luminous object.

This research is based on observations with the NASA/ESA Hubble Space Telescope, obtained at the STScI, which is operated by the AURA under NASA contract No. NAS 5-26555. Authors acknowledge financial support by NASA STI 9377.05-A. Ueta also acknowledge partial support by a US National Research Council Research Associateship Award and a NASA Postdoctoral Program Research Fellowship Award. An anonymous referee is thanked by the authors for his/her comments.

TABLE 1
*HST*/NICMOS Imaging-Polarimetry Observing Log of the Egg

| Date | FILTER[a] | SAMP_SEQ | NSAMP | DITHER PATTERN | NPTS | EXPTIME[b] (sec) | ORIENTAT[c] (deg) | PID | Reference |
|---|---|---|---|---|---|---|---|---|---|
| 1997 Apr 8 | POL-L | STEP16 | 26 | SPIRAL | 4 | 1215.39 | 56.3451 | 7115 | 1, 2, 3, 4 |
| 2002 Oct 6 | POL-L | STEP16 | 26 | SPIRAL | 4 | 1215.39 | −117.752 | 9644 | 4 |

References. — 1. Sahai et al. (1998), 2. Hines et al. (2000), 3. Weintraub et al. (2000), 4. Ueta et al. (2005b)

[a] POL-L: long wavelength polarizers ($1.89 - 2.1 \mu$m, centered at $2.05 \mu$m).
[b] Total exposure time per polarizer.
[c] The ORIENTAT header parameter refers to PA of the image +y axis (degrees E of N).

TABLE 2
NICMOS POL-L Plate Scales

| Epoch (Date) | X (arcsec pixel$^{-1}$) | Y (arcsec pixel$^{-1}$) |
|---|---|---|
| 1 (1997 Apr 8) | 0.07645 | 0.07576 |
| 2 (2002 Oct 6) | 0.07595 | 0.07536 |

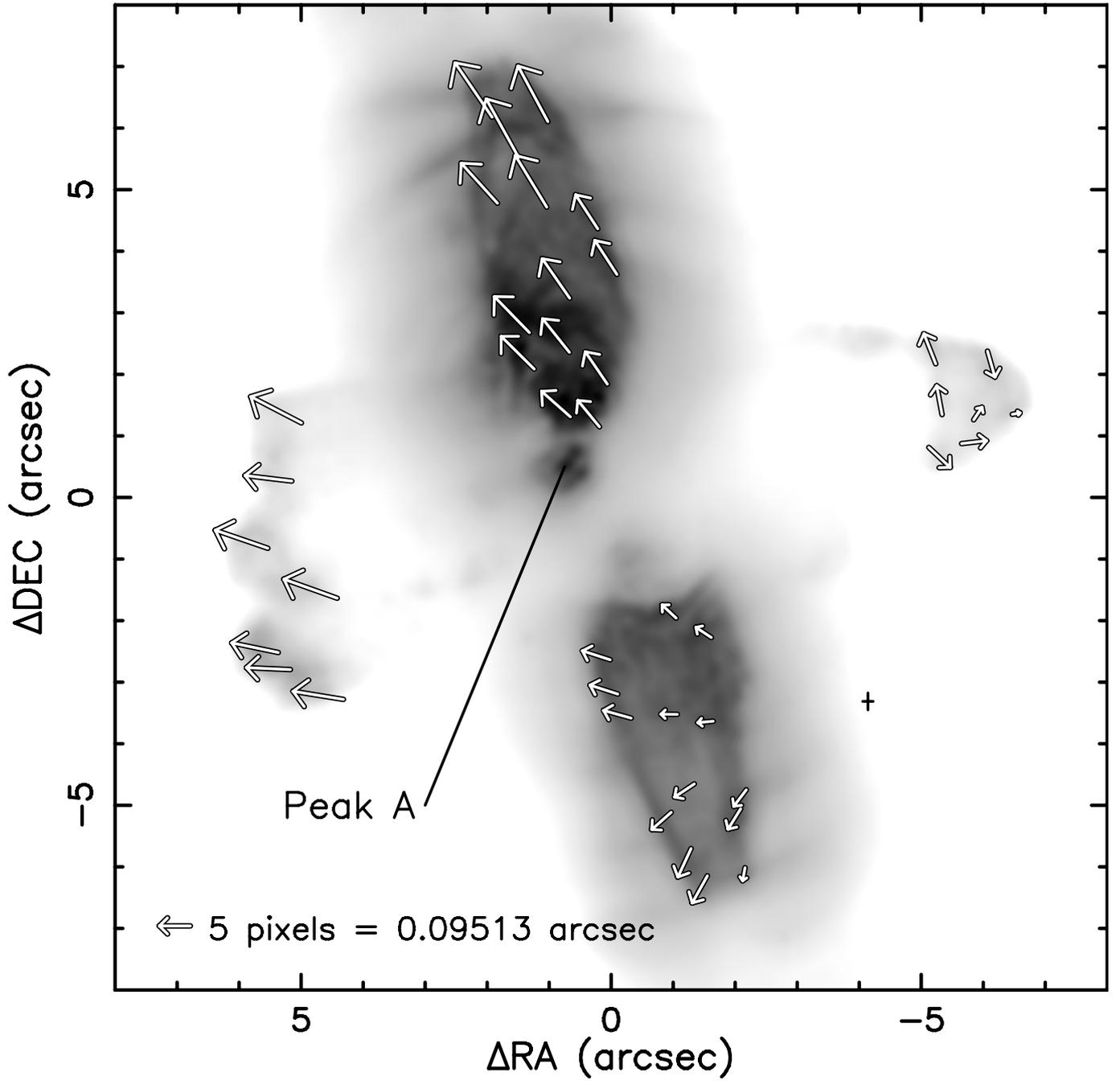

FIG. 1.— The grayscale Stokes *I* image of the Egg (N is up, E to the left) at the POL-L (2$\mu$m) band overlaid with proper-motion vectors of the selected local structures. The vectors, placed at the position of the structures, depict the direction and amount of motion in the past 5.5 years. A vector corresponding to 5-pixel shift (which equals to 0.″09513, i.e. 17.3 mas yr$^{-1}$; the scale for the vector length is not equal to the image scale) is displayed on the lower left corner for reference. The plus sign shows the origin of the outflow backtracked from these vectors, which is obviously off the nebula. The size of the plus sign indicates the accuracy of the origin determination.

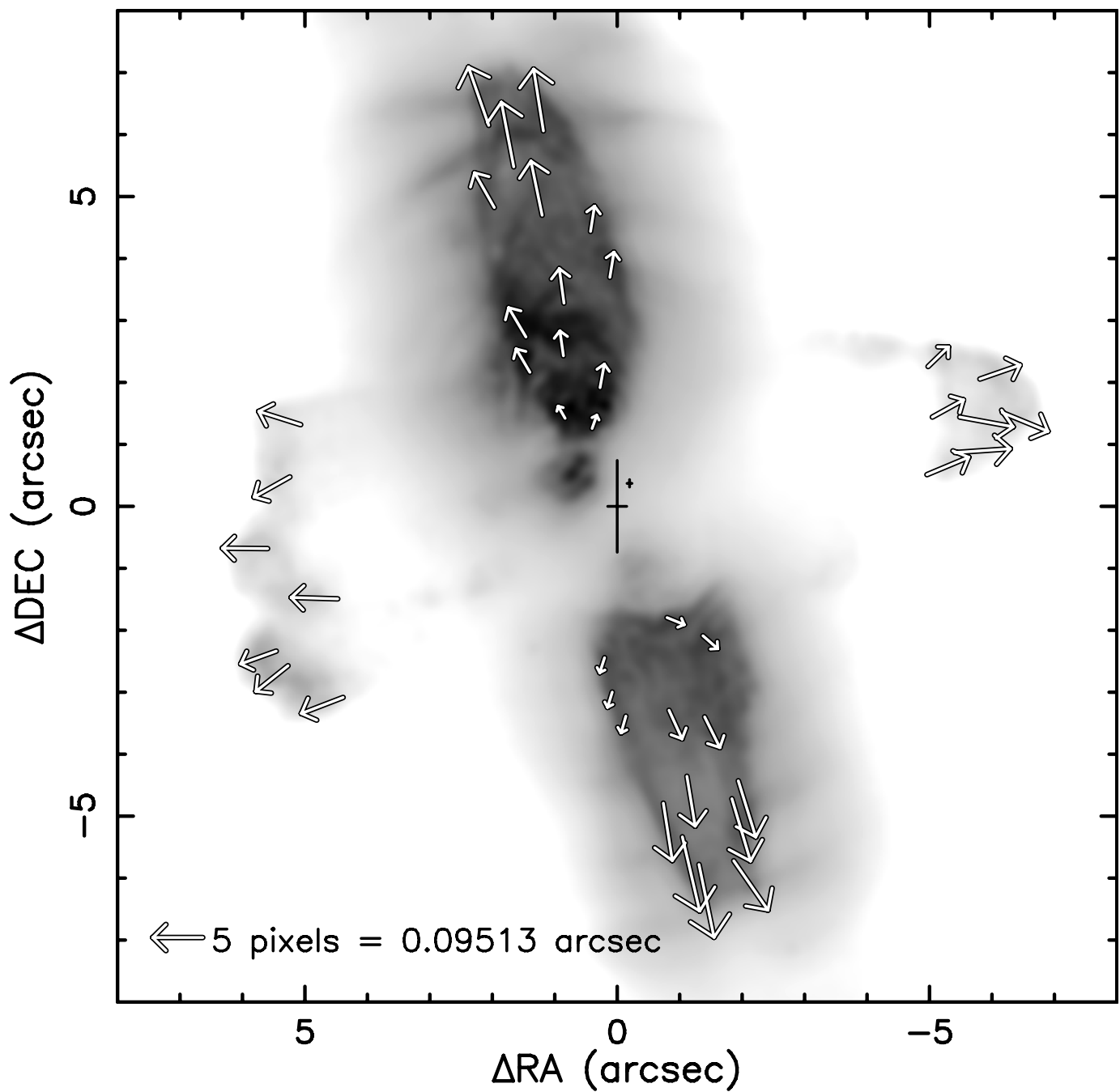

FIG. 2.— Same as Figure 1, but the vectors are obtained based on the the epoch 2 image that is corrected for the stellar proper motion of the Egg itself. The larger plus sign at (0, 0) indicates the origin of the outflow backtracked from these vectors, while the smaller plus sign at (−0.2, 0.4) marks the origin of the polarization vector pattern derived by (Weintraub et al. 2000). The size of the signs shows the accuracy of the origin determination in respective analyses (enlarged by a factor of 2 for clarity).

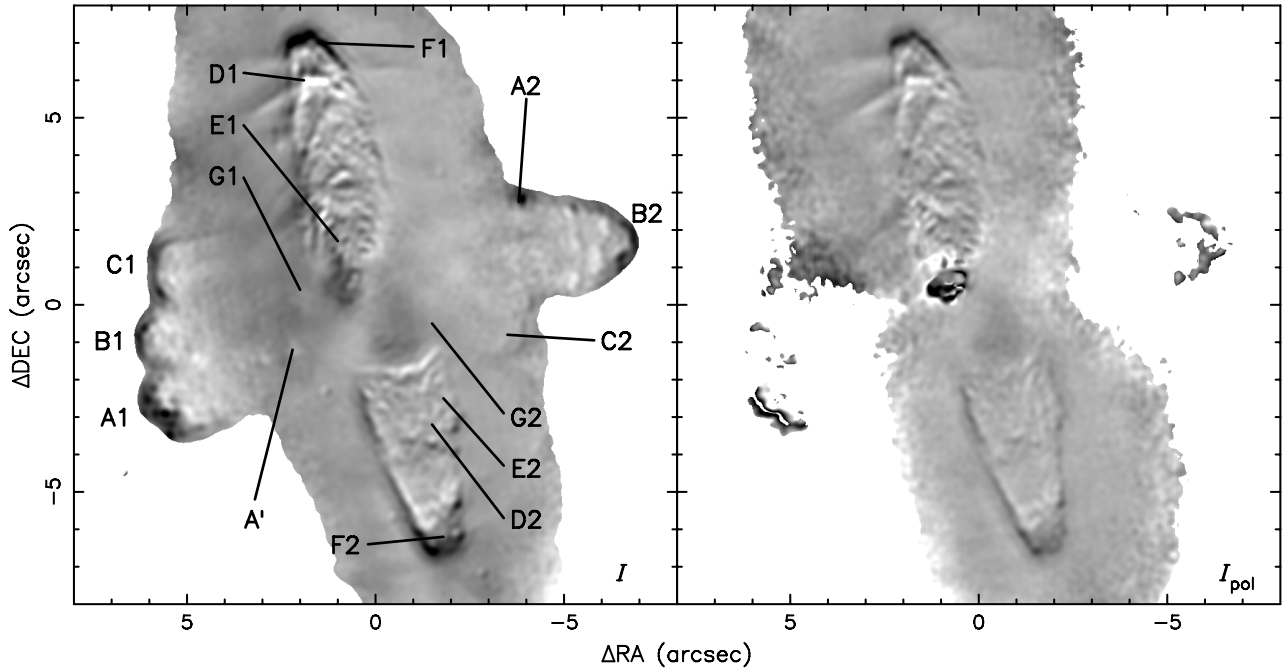

FIG. 3.— The epoch 2 over epoch 1 ratio images in the Stokes $I$ (left) and polarized light (right). The darker the color becomes the more emission there is in the epoch 2 image. The A1/A2 through G1/G2 designation corresponds to the tips of CO J=2-1 outflows detected by Cox et al. (2000).

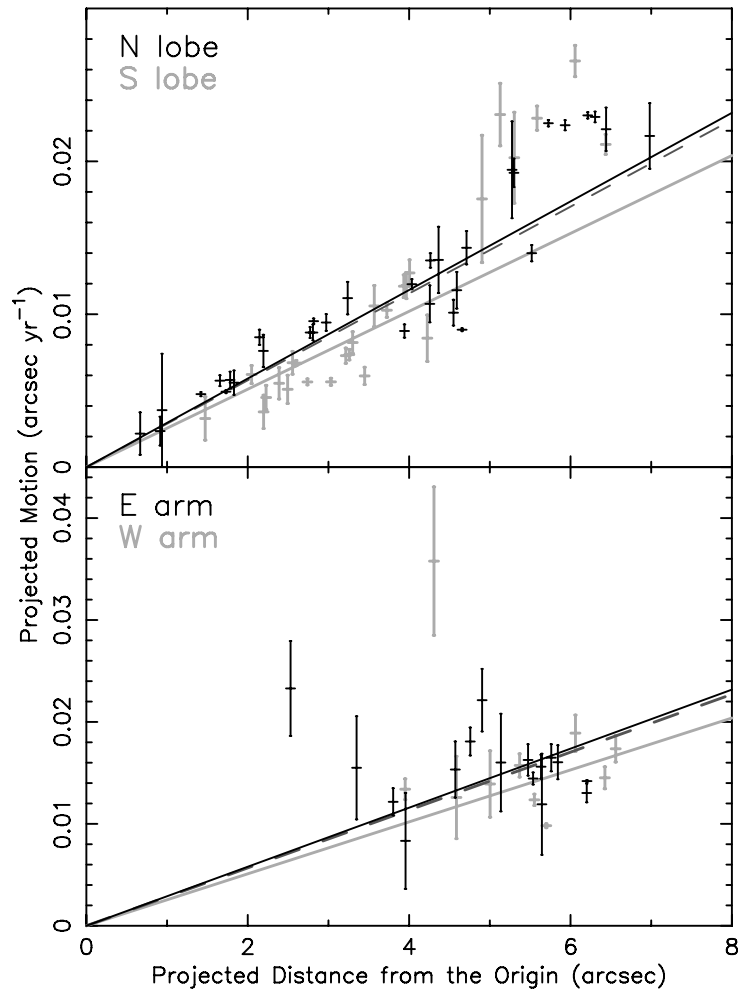

FIG. 4.— Proper motion of nebula (arcsec per year) as a function of projected radial distance from the origin of the expansion (arcsec). [Top] Motion of local structures in the north lobe (black) and in the south lobe (light gray) with error bars. The least-squares fit to the points are also shown (solid black - N lobe; solid light gray - S lobe; dashed dark gray - both lobes). [Bottom] The same as the top frame, but for structures in the east arm (black) and west arm (light gray). The least-squares fit to the points in the N & S lobes are duplicated for reference.